\documentclass[prl,twocolumn,showpacs,amsmath,amssymb,superscriptaddress]{revtex4}

\usepackage{graphicx}
\usepackage{dcolumn}
\usepackage{bm}
\usepackage{latexsym}

\newcommand{\bmat}{\begin{displaymath}}

\newcommand{\emat}{\end{displaymath}}

\newcommand{\bit}{\begin{itemize}}

\newcommand{\eit}{\end{itemize}}

\newcommand{\beq}{\begin{equation}}

\newcommand{\eeq}{\end{equation}}

\newcommand{\bspl}{\begin{split}}

\newcommand{\espl}{\end{split}}

\begin{document}

\preprint{}

\title{Electron Interactions and Transport Between Coupled Quantum Hall 
Edge States}

\author{J. W. Tomlinson} 
\affiliation{Theoretical Physics, University of Oxford, 1 Keble
Road, OX1 3NP, United Kingdom.}
\author{J.-S. Caux} 
\affiliation{Institute for Theoretical Physics, University of
  Amsterdam, Valckenierstraat 65, 1018 XE Amsterdam, The Netherlands.}
\author{J. T. Chalker} 
\affiliation{Theoretical Physics, University of Oxford, 1 Keble Road,
  OX1 3NP, United Kingdom.}

\date{\today}

\begin{abstract}
We examine the effects of electron-electron interactions on transport between
edge states in a multilayer integer quantum Hall system. 
The edge states of such a system, coupled by interlayer tunneling, form a two-dimensional, chiral metal at 
the sample surface. We calculate the temperature-dependent conductivity and the amplitude
of conductance fluctuations in this chiral metal, treating Coulomb interactions
and disorder exactly in the weak-tunneling limit. We find that the conductivity increases
with increasing temperature, as observed in recent experiments, and we show that the correlation
length characterising conductance fluctuations varies inversely with temperature.
\end{abstract}





\maketitle
%
%
Layered conductors in magnetic fields are interesting partly because
they can show a three-dimensional version of the quantum Hall effect
if interlayer coupling is weak. In these circumstances, with a magnetic
field perpendicular to the layers, the Hall conductance
of individual layers is quantised and edge states are present in each layer at the sample surface.
Interlayer tunneling couples edge states from adjacent layers to form 
a surface phase, which is a chiral, two-dimensional metal. 
The surface phase is predicted \cite{dohmen,balents} and observed \cite{gwinn,otherexpt}
to dominate transport along the interlayer direction in small samples at low temperature,
since, for a system within a quantum Hall plateau, states in the bulk are
Anderson localised. 

The transport properties of the chiral metal are different in striking ways 
from those of other two-dimensional conductors. In particular, even in the presence
of strong disorder, chiral motion of electrons around the sample perimeter acts 
to suppress localisation of the surface states \cite{dohmen,balents}.
Experimentally, this is demonstrated by conduction which is metallic, in the 
sense that the surface conductivity remains non-zero in the low-temperature limit,
even for samples in which this limiting value
is very much smaller than $e^2/h$ \cite{gwinn}. Although scattering by disorder
does not lead to localisation, it is 
expected theoretically to show itself via
the existence of mesoscopic conductance fluctuations \cite{mathur,gruzberg,cho,plerou,betouras}.
Observations made of these fluctuations \cite{flucs} help to
confirm that electron motion between 
edges is indeed via phase-coherent tunneling, rather than by incoherent hopping.

The consequences of electron interactions for the chiral 
metal have so far
attracted much less attention than the effects of 
impurity scattering outlined above.
In this connection, the recent 
discovery that, unusually for a metal, the conductivity $\sigma(T)$ 
{\it increases}
with increasing temperature $T$ \cite{tempdep} is very striking.
Two straightforward potential reasons for such behaviour
are excluded by the experimental design and analysis of Ref.\cite{tempdep}.
It cannot be due to weak localisation
effects which arise when electrons circumnavigate the
sample without losing phase coherence, since sample perimeters
are much longer than the inelastic scattering length.
And it cannot be due to transport through
the bulk of the sample, because that is shown to be 
negligible in the temperature range
range (50 - 400 mK) of interest. Electron interactions stand out as
the likely source of $T$-dependence in $\sigma(T)$
and this provides one of the motivations for 
the work described here. A second motivation stems from the fact that, in the experimentally-relevant limit
of weak interlayer tunneling, a set of coupled integer quantum Hall edge states
offers a rare example of a system in which both disorder and Coulomb interactions can be
treated exactly.

In the following, we obtain for the chiral metal the full temperature dependence of the conductivity
and the autocorrelation function of conductance fluctuations, working at leading order
in interlayer tunneling strength. Calculations depend on the fact that, without
tunneling, the only excitations of the system are collective harmonic modes -- surface magnetoplasmons --
which can be treated straightforwardly using bosonization. Coulomb interactions
introduce dispersion into the spectrum of these excitations, which in turn is responsible
for temperature dependence of transport properties. 
With reasonable assumptions, discussed below, our results for the variation of
$\sigma(T)$ with $T$ are consistent with experiment \cite{tempdep}. 

The physical ingredients that are important for our 
results contrast in two
obvious ways with what is stressed in work on fractional edge states,
including the treatment of multilayer fractional 
quantum Hall systems in Ref.~\cite{naud}.
First, the integer edge states considered here are intrinsically simpler than 
fractional edge states, in that they are Fermi rather than Luttinger liquids.
Second, and conversely, the temperature-dependence we find requires a
proper treatment of Coulomb interactions and would not arise with 
the contact interactions implicit in most treatments of fractional edge states.

Turning to calculations,
our starting point is a Hamiltonian for the surface states,
which decouple from the bulk states because the latter are localised:
 ${\cal H} = {\cal H}_{\rm 0} + {\cal H}_{\rm int} 
+ {\cal H}_{\rm hop}$. 
It has a single-particle contribution  ${\cal H}_{\rm 0}$ for uncoupled layers,
each with one edge state and
impurity scattering, an interaction term ${\cal H}_{\rm int}$, 
and interlayer hopping ${\cal H}_{\rm hop}$.
We denote the (bare) edge velocity by $v$ and the interlayer hopping amplitude by $t_{\perp}$,
take the impurity potential at position $x$ on the edge of the $n$th layer to be $V_n(x)$
and write the interaction potential as $U_{n}(x)$. 
Introducing the electron creation operator $\psi^{\dagger}_n(x)$ 
and density $\rho_n(x)=\psi^{\dagger}_n(x)\psi^{\phantom{\dagger}}_n(x)$,
with normalisation fixed by $\{\psi^{\dagger}_n(x),\psi^{\phantom{\dagger}}_m(x')\}=\delta_{nm}\delta(x-x')$,
the terms in $\cal H$ are
\begin{eqnarray}
{\cal H}_0 = \sum_n \int dx \psi^{\dagger}_n(x)[-{\rm i}\hbar v\partial_x + V_n(x)]\psi^{\phantom{\dagger}}_n(x) \nonumber \\
{\cal H}_{\rm int} = \frac{1}{2}\sum_{nm}\int dx \int dx' \rho_n(x) U_{n-m}(x-x') \rho_m(x') \nonumber \\
{\cal H}_{\rm hop} = \sum_n \int dx [t_{\perp}\psi^{\dagger}_{n+1}(x)\psi^{\phantom{\dagger}}_n(x)+ 
{\rm h.\,c.}]\,.
\label{hamiltonian}
\end{eqnarray}

The non-interacting system, ${\cal H}_{\rm 0} + {\cal H}_{\rm hop}$, can be characterised by three lengthscales,
$l_{\rm el}$, $l_{\perp}$ and $L_{\rm T}$, as follows.
First, choosing a Gaussian impurity potential distribution
with $\langle V_n(x) V_m(x') \rangle = \Delta \delta_{nm}\delta(x-x')$, the elastic mean free path
for forward scattering within a layer is $l_{\rm el}= (\hbar v)^2/\Delta$ \cite{balents}.
Second, denoting layer spacing by $a$, the diffusion constant for interlayer motion is 
$D=a^2v/l_{\perp}$, where $l_{\perp} = \Delta/2t_{\perp}^2$ can be interpreted as the distance traveled
in the chiral direction between interlayer transitions \cite{betouras}. Third, temperature $T$
can be expressed in terms of a thermal length $L_{\rm T} = \hbar v/k_{\rm B} T$. We are concerned
with the regime $l_{\rm el} \ll L_{\rm T} \ll l_{\perp}$.

For Coulomb interactions with layer separation $a$
\beq
U_n(x) = \frac{e^2}{4\pi \epsilon_r \epsilon_0 \sqrt{x^2 + n^2a^2 + w^2}}\,,
\label{Coulomb}
\eeq
where the parameter $w$ has been introduced to model the finite width of an edge state.
Interactions lead to a renormalisation
of the edge velocity, from $v$ to $v_{\rm F}$, and we redefine $L_{\rm T}$ accordingly.
Beyond this, 
the interaction strength is characterised
by the inverse screening length $\kappa  = e^2/4\pi \epsilon_r \epsilon_0 \hbar v_{\rm F} a$:
for current experiments \cite{gwinn,flucs,tempdep} we estimate below that $\kappa a \ge 1$, depending on
assumptions about $v_{\rm F}$.

In outline (details will be presented elsewhere \cite{long}), our calculations center on
the Kubo formula
\beq
\sigma(T)=\frac{{\rm i}a}{\hbar}\sum_m \int dx \int_{-\infty}^{\infty} \!\!\! dt\,t \,
[\langle j_m(x,t) j_n(0,0)\rangle]_{\rm av}\,,
\label{kubo}
\eeq
where $\langle \ldots \rangle$ and $[\ldots ]_{\rm av}$ denote thermal and disorder averages, respectively,
the Heisenberg representation ${\cal O}(t) \equiv  {\rm e}^{{\rm i} {\cal H}t/\hbar} {\cal O} {\rm e}^{-{\rm i} {\cal H}t/\hbar}$
is used, and the current operator is
\beq
j_n(x) = \frac{{\rm i} e }{\hbar} [t_{\perp} \psi^{\dagger}_{n+1}(x)\psi^{\phantom{\dagger}}_n(x)-
{\rm h.\,c.}]\,.
\label{current}
\eeq
In the weak-tunneling limit it is sufficient to retain $t_{\perp}$ only in $j_n(x)$ and to evaluate
the thermal average omitting ${\cal H}_{\rm hop}$ from $\cal H$; since $t_{\perp}$ is a relevant perturbation \cite{naud}
our approach is justified for temperatures that are not too low: $L_{\rm T} \ll l_{\perp}$.
Moreover, a gauge transformation can be used to remove $V_n(x)$ from ${\cal H}_0$,
transferring it instead to ${\cal H}_{\rm hop}$ and $j_n(x)$ by means of the replacements in Eqns.\,(\ref{hamiltonian}) and (\ref{current}):
$\psi_n(x) \rightarrow e^{-{\rm i} \theta_n(x)} \psi_n(x)$
and
$t_{\perp} \rightarrow t_{\perp}(n,x) \equiv t_{\perp} e^{{\rm i}[\theta_{n+1}(x) - \theta_{n}(x)]}$
with $\hbar v \partial_x \theta_n(x) =  V_n(x)$ \cite{bdycond}.
The conductance and its fluctuations at small $t_{\perp}$ can therefore be expressed in terms of
the electron correlation function without disorder evaluated at $t_{\perp}=0$,
\beq
G(x,t) = \langle \psi^{\dagger}_{n}(x,t)\psi^{\phantom{\dagger}}_{n+1}(x,t)\psi^{\dagger}_{n+1}(0,0)\psi^{\phantom{\dagger}}_n(0,0)\rangle\,,
\eeq
combined with the disorder average of products of $t_{\perp}(n,x)$ and $t^*_{\perp}(n,x)$.

Proceeding with the calculation of $\sigma(T)$,
we use the result $\langle t_{\perp}(n,x) t_{\perp}(m,x')\rangle =
\delta_{nm}
\exp(-|x-x'|/l_{\rm el})$ to write Eq.\,(\ref{kubo}) as
\beq
\sigma(T) = -\frac{2 a}{\hbar} ( \frac{ e t_{\perp}}{\hbar})^2 \int_{-\infty}^{\infty} 
\! \! dx\, {\rm e}^{-|x|/l_{\rm el}} \int_{-\infty}^{\infty}\!\! dt\, t\, {\rm Im}G(x,t)
\label{kubo2}
\eeq
and for $l_{\rm el} \ll L_{\rm T}$ simplify it to
\beq
\sigma(T) = -\frac{4 a l_{\rm el}}{\hbar} ( \frac{ e t_{\perp}}{\hbar})^2 \int_{-\infty}^{\infty}\!\! dt\, t\, {\rm Im}G(0,t)\,.
\label{kubo3}
\eeq

We compute $G(x,t)$ by switching to a bosonic description of the system in the standard way,
introducing a field $\phi_n(x)$ in each layer, related
to the electron density by $\rho_n(x) = (2\pi)^{-1}\partial_x \phi_n(x)$ and to
the electron field by $\psi^{\dagger}_n(x) = (2\pi \epsilon)^{-1/2} {\rm e}^{{\rm i}\phi_n(x)}$,
where $\epsilon$ is a short-distance cut-off \cite{klein}. It has the mode expansion
\beq
\phi_n(x) = -\sum_q (\frac{2\pi}{Lq})^{1/2}({\rm e}^{{\rm i}qx} b^{\dagger}_{qn} + {\rm h.\,c.}){\rm e}^{-\epsilon q/2}\,,
\eeq
where $[b^{\dagger}_{q'm},b^{\phantom{\dagger}}_{qn}]=-\delta_{qq'} \delta_{nm}$,
$q=2\pi n_q/L$ for a system of circumference $L$ in the chiral direction, and $n_q$ is a positive integer.
The Hamiltonian omitting hopping is quadratic in the boson fields:
it is diagonalised by Fourier transform in the interlayer direction. Setting
$b_{qk}=N^{-1/2}\sum_n{\rm e}^{-{\rm i}nka}b_{qn}$ for an $N$-layer system
with periodic boundary conditions in both directions, where $k=2\pi n_k/Na$ and $0 \leq n_k < N$ is integer, we have
\beq
{\cal H}_0 + {\cal H}_{\rm int} = \sum_{qk} \hbar \omega(q,k) b^{\dagger}_{qk} b^{\phantom{\dagger}}_{qk}\,.
\eeq
The collective excitation frequencies $\omega(q,k)$ appearing here 
depend on the edge velocity and the interaction potential.
From the Fourier transform of $U_n(x)$ we define the wavevector-dependent velocity
\beq
u(q,k)=(2\pi \hbar)^{-1}\sum_n\int dx\,  {\rm e}^{{\rm i}(qx+nka)}U_n(x)\,.
\label{int}
\eeq
The renormalised edge velocity is
$v_{\rm F}= v- u(0,0)$, where contributions to $v$ from a neutralising background
cancel the divergence in $u(0,0)$. The excitation frequencies are
$\omega(q,k) = [v_{\rm F} + u(q,k)]q$.
For Coulomb interactions, from Eq.\,(\ref{Coulomb}), we have
\beq
u(q,k) = v_{\rm F} \sum_p \frac{\kappa}{Q_p} {\rm e}^{-wQ_p}
\eeq
with $Q_p^2 = q^2 + (k+2\pi p/a)^2$ and $p$ integer. In the continuum
limit ($a \rightarrow 0$) at small $w$, the dispersion relation simplifies to
\beq
\omega(q,k) = v_{\rm F}q(1 + \frac{\kappa}{\sqrt{q^2 + k^2}})\,.
\label{continuum}
\eeq
This result can be contrasted wih the dispersion relation
for edge magnetoplasmons in a single-layer quantum Hall
system \cite{volkov}: $\omega(q) \propto q \ln(1/qw)$.

The correlation function we require can be expressed as an integral over
excitation modes: defining $S$ via $G(x,t)=(2\pi)^{-2}{\rm e}^S$, we find \cite{long}
\begin{equation}
\begin{split}
&S=-2\log{\epsilon}-\frac{a}{\pi}\int^{\pi/a}_{-\pi/a}dk(1-\cos{ka})\!\!\int_0^{\infty}
\frac{dq}{q}e^{-\epsilon q}\\
&\!\times\!\!\bigg(\!\!\coth{\!\left(\!{\beta\hbar\omega({q,k})}/{2}\!\right)}\!
\left[1\!-\!\cos{(qx +\omega({q,k})t)}\right]\\
&\qquad\qquad\qquad\qquad\qquad+{\rm i}\sin(qx+\omega(q,k)t)\bigg)\label{eq:fullS}.
\end{split}
\end{equation}
Combining Eqns.\,(\ref{kubo3}) and (\ref{eq:fullS}),
and evaluating the integrals on $k$, $q$ and $t$ numerically, we obtain $\sigma(T)$.

\begin{figure}
\begin{center}
\includegraphics[width=0.40\textwidth]{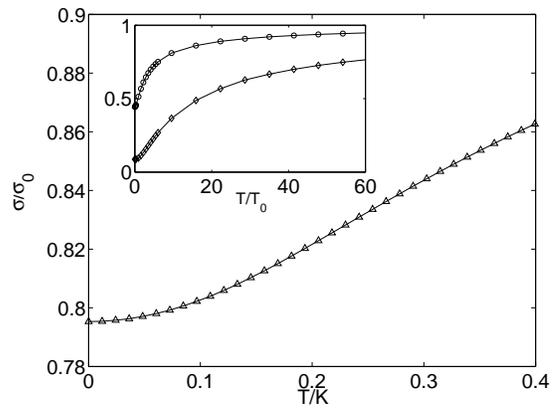}
\caption{\label{fig1} Conductivity vs.
temperature, calculated for
wide edge state with parameters chosen to reproduce experimental
results
of Ref.\cite{tempdep}.
Inset: results for narrow edge states with interaction
strengths $\kappa a=1$ ($\circ$) and $\kappa a=5$ ($\diamond$).}
\end{center}
\vspace{-0.7cm}
\end{figure}

To discuss the results, it is useful to start from the
expression for conductivity with interactions omitted except in the value of $v_{\rm F}$,
$\sigma_0 = (e^2/h)\cdot 2a l_{\rm el} \cdot (t_{\perp}/\hbar v_{\rm F})^2$ \cite{balents,betouras}.
The form that this takes can be understood in terms of a calculation
of the tunneling rate between adjacent edges, 
where $v_{\rm F}^{-2}$ enters through a product of the densities
of initial and final states, since the density of states per unit length
and energy for a single edge is $(2\pi \hbar v_{\rm F})^{-1}$.
To account fully for interactions, one should, in essence, replace $v_{\rm F}$ in
this expression by the group velocity $\partial_q \omega(q,k)$ and perform an
appropriate thermal average. Since the group velocity
is maximum at $\omega(q,k)=0$ and decreases with increasing
$\omega(q,k)$, the conductivity is minimum at $T=0$ and increases with $T$.
These features are apparent in the inset to Fig.\,\ref{fig1}, which shows
$\sigma(T)$ calculated using the dispersion 
relation for small $w$, Eq.\,(\ref{continuum}), with temperature measured in units of
$T_0=\hbar v_{\rm F}/a k_{\rm B}$ (so that $T/T_0 = a/L_{\rm T}$).
In particular, $\sigma(0)$ is suppressed relative to its value at $\kappa = 0$,
by a factor that increases with increasing $\kappa$,
while $\sigma(\infty)=\sigma_0$, independently of $\kappa$. At low temperatures, 
$\sigma(T) - \sigma(0) \propto T^2$.

The experiments of Ref.\,\cite{tempdep} share these qualitative features,
but a quantitative match appears to require refinement of our model,
which we now describe.
Specifically, an approximately linear variation of $\sigma(T)$ against $T$
is reported \cite{tempdep}, with an increase of $\sim 8\%$ over $50{\rm mK} \leq T \leq 300{\rm mK}$.
(These data are in fact for filling factor per layer close to $\nu=2$,
so that 
two edge states are present: we assume that
they behave independently and can each be modeled by Eq.\,(\ref{hamiltonian}).
In other respects, the samples appear to be in the regime treated by our theory:
$l_{\rm el} \approx 30{\rm nm}$ \cite{magnetoresistance} and $l_{\perp} \approx 40\mu{\rm m}$ (from the magnitude of $\sigma(T)$),
so $l_{\rm el} \ll l_{\perp}$.)
As a fitting parameter, we have the value of $v_{\rm F}$, which enters $\kappa$ and $T_0$.
An upper bound on $v_{\rm F}$, reached in samples with a steep confining potential for electrons at the surface,
is $v_{\rm F} \sim \omega_{\rm c} l_{\rm B}$, where $\omega_{\rm c}$ and $l_{\rm B}$ are
the cyclotron frequency and magnetic length.
Under experimental conditions (GaAs at 6.75 tesla, with $a=30{\rm nm}$), $\omega_{\rm c} l_{\rm B} \approx 1.7 \times 10^5 {\rm ms}^{-1}$,
which implies $\kappa a\sim 1$ and $T_0 \sim 40\, {\rm K}$. With these values, the
variation of $\sigma(T)$ over the experimental range of $T$ would be {\it very small} and
{\it quadratic}, in disagreement with observations \cite{tempdep}. To fit experiment, we require  
a reduction in the temperature scale on which $\sigma(T)$ varies. This is 
ensured by a smaller value for
$v_{\rm F}$ and by a finite width $w$ for edge states.
For example, setting $v_{\rm F}=3\times 10^3\,{\rm ms}^{-1}$, so that $\kappa a=50$
and $L_{\rm T}=200{\rm nm}$ at $100{\rm mK}$, and
using an edge state width $w=120\,{\rm nm}$,
we obtain the results shown 
for $0\leq T \leq 400{\rm mK}$ in Fig.\,\ref{fig1}, which are close to those of Fig.~4 in Ref.\,\cite{tempdep}.

There are in fact separate grounds for expecting $w$ to be of about this size. From a theoretical viewpoint,
while in clean samples $w\sim l_{\rm B} \sim 10\,{\rm nm}$, more generally one expects $w \sim \xi$, where $\xi$ is the
localisation length for bulk states. From experiment, first, 
an analysis of bulk hopping transport (which dominates over surface transport at higher temperatures)
yields $\xi \sim 120 {\rm nm}$ \cite{hopping}. Second, in studies of
conductance fluctuations in the chiral metal, induced by small variations in magnetic field within a quantum Hall
plateau, comparison of the fluctuation amplitude with the correlation field
suggests $w \sim 70 {\rm nm}$ \cite{flucs}. 
A large value of $w$ in turn favours a small value for $v_{\rm F}$,
since wide edge states extend into the bulk of the sample where the
confining potential gradient is small.
Moreover, while the small value we have used for $v_{\rm F}$ increases $\sigma_0$,
a large value for $w$ has the opposite effect of reducing the effective matrix element between edges:
such a compensation is necessary to account correctly for the absolute magnitude of the 
experimental $\sigma(T)$ \cite{long}.

\begin{figure}[t]
\begin{center}
\includegraphics[width=0.40\textwidth]{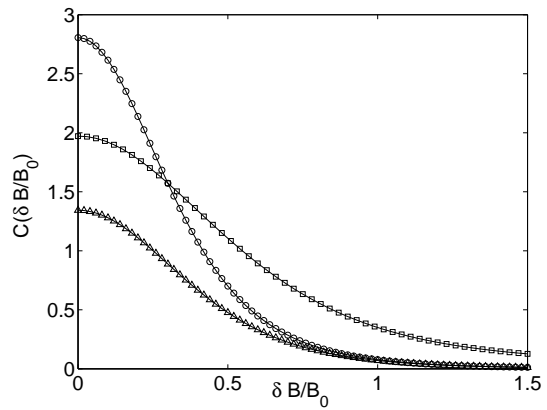}
\caption{\label{fig3} The scaling function $C(\delta B/B_0)$ for conductance fluctuations:
for narrow edges with $\kappa a=0.6$ ($\circ$) and $\kappa a=1$ ($\triangle$); and for wide edges with the
parameters of Fig.\,\ref{fig1} ($\Box$).}
\end{center}
\vspace{-0.7cm}
\end{figure}

We turn finally to a theoretical treatment of fluctuations $\delta g(B_{\perp})$ in the conductance
of a finite sample. For simplicity, we suppose these fluctuations are induced by varying the magnetic field component 
$B_{\perp}$ normal to the plane of the chiral metal, although in experiment it is
the magnetic field normal to the layers which is varied, with finite $w$ probably responsible
for sensitivity of the conductance to this field component. In general, the amplitude of
mesoscopic conductance fluctuations may be limited either by inelastic scattering or by
thermal smearing. However, in a chiral metal without interactions, states are perfectly
correlated in energy (single-particle eigenfunctions of ${\cal H}_0 + {\cal H}_{\rm hop}$
with distinct energies differ only by a factor ${\rm e}^{{\rm i} qx}$) and so thermal smearing
is absent \cite{betouras}. We are therefore concerned solely with
interaction effects.
Our aim is to compute
$F(\delta B) \equiv \langle \delta g(B_{\perp}) \delta g(B_{\perp} + \delta B)\rangle$.
Defining $g_0=\sigma_0 L/N$, the average conductance at $\kappa =0$,
and $b = \delta B/B_0 = \delta B a L_{\rm T}e/ \hbar$, the number of flux quanta 
threading a rectangle of area $2\pi \times L_{\rm T} \times a$, we obtain \cite{long} at 
leading order in $t_{\perp}$, with $l_{\rm el} \ll L_{\rm T}$, 
the expression
\beq
F(\delta B) = \frac{g_0^2}{NL} \int_{-\infty}^{\infty} dx {\rm e}^{{\rm i} bx/L_{\rm T}}\left[4\pi v_{\rm F}^2
\int_{-\infty}^{\infty}\!\! dt\, t\, {\rm Im} G(x,t)\right]^2\,.
\nonumber
\eeq
In the low-temperature regime, where $\sigma(T) \approx \sigma(0)$, this 
has the scaling form
\beq
F(\delta B) = \frac{g_0^2 L_{\rm T}}{NL} C(\delta B/B_0)\,,
\eeq
where temperature $T$ and magnetic field difference $\delta B$ enter only
through the scaling variables $L_{\rm T}/L$ and $\delta B/B_0$, but the scaling
function $C(\delta B/B_0)$ varies with the interaction potential. It is illustrated in Fig.\,\ref{fig3},
calculated both using the dispersion relation for small $a$ and $w$, Eq.\,(\ref{continuum}),
and for the parameters of Fig.\,\ref{fig1}. 
An inelastic scattering length can be identified either from
the amplitude of conductance fluctuations or from their correlation field.
By either route, it is proportional to $L_{\rm T}$ and hence varies as $T^{-1}$.
It also depends on interaction strength, because of the variation of $C(\delta B/B_0)$:
an increase in the inelastic scattering length with decreasing $\kappa$
is reflected in an increased amplitude and decreased width of $C(\delta B/B_0)$.
Such a dependence of the inelastic scattering length
on interaction strength and temperature is long-established in
non-chiral one-dimensional conductors \cite{apel}.
For the chiral metal, this $T^{-1}$ dependence is as conjectured previously \cite{balents}
for weakly coupled edges, and stands in contrast to the $T^{-3/2}$ behaviour,
obtained in perturbation theory for strongly coupled edges \cite{betouras}.
%
%
With values for $\kappa$ and $w$
chosen to reproduce the observed $\sigma(T)$, the calculated amplitude of conductance fluctuations
is about 60\% of the measured one \cite{flucs}, which we regard as adequate agreement.


We thank E. G. Gwinn for very helpful discussions and J. J. Betouras
for previous collaborations. The work was supported by EPSRC under Grant GR/R83712/01
and by  the Dutch FOM foundation. 

\vspace{-0.6cm}
%
%


\begin{thebibliography}{99}
\bibitem{bdycond} We neglect the effect of this gauge transformation on
boundary conditions, which is justfied for $L_{\rm T} \ll L$.
\bibitem{klein} We omit Klein factors, which cancel from our final expressions.
\end{thebibliography}
\end{document}